%% file: 2003Puebla.tex
\documentclass{Rinton-P9x6}
\usepackage{latexsym}
\usepackage{epsfig}

\def\H{{\cal H}}
\def\L{{\cal L}}

\def\op#1{\hat{#1}}
\def\ket#1{| #1 \rangle}
\def\lket#1{| #1 \rangle\rangle}
\def\bra#1{\langle #1 |}

\def\vec#1{{\bf #1}}

\def\Tr{{\rm Tr}}

\def\rmi{{\rm i}}

\newif\ifpdflatex\pdflatextrue
\makeatletter\@ifundefined{pdfoutput}{\pdflatexfalse}\makeatother
\def\myincludegraphics[#1]#2#3{%
\ifpdflatex \includegraphics[#1]{#2}
\else       \includegraphics[#1]{#3}
\fi}
\begin{document}
\title{Dissipative Quantum Control\footnote{Talk presented at \emph{ICCSUR 8}, Puebla, Mexico, July 2003}}
\author{Allan I.\ Solomon$^{1,2}$ and Sonia G.\ Schirmer$^{3}$}
\address{${}^1$ Department of Physics and Astronomy, The Open University, \\
                Milton Keynes, MK7 6AA, United Kingdom \\
         ${}^2$ LPTL, University of Paris VI, France\\
         ${}^3$ Dept of Applied Maths \& Theoretical Physics,
                University of Cambridge, \\ Wilberforce Road, 
                Cambridge, CB3 0WA, United Kingdom\\
         E-mail: a.i.Solomon@open.ac.uk, sgs29@cam.ac.uk}
\maketitle
\abstracts{Nature, in the form of dissipation, inevitably intervenes in our 
efforts to control a quantum system.  In this talk we show that although we 
cannot, in general, compensate for dissipation by coherent control of the 
system, such effects are not always counterproductive; for example, the 
transformation from a thermal (mixed) state to a cold condensed (pure state)
can only be achieved by non-unitary effects such as population and phase
relaxation.}

\section{Representation of quantum states for dissipative systems}

In closed-system, pure-state quantum mechanics the state of the system is usually
represented by a wavefunction $\ket{\Psi}$, which is an element of the Hilbert 
space $\H$.  For open quantum systems, however, a quantum statistical mechanics 
formulation is necessary since dissipative effects due to the interaction of the
system with its environment convert pure states into statistical ensembles and 
vice versa.  The state of the system must therefore be represented by a density
operator $\op{\rho}$, i.e., a positive trace-one operator acting on $\H$.  It 
is convenient to expand this density operator in terms of a complete orthonormal
set of energy eigenstates $\{\ket{n}:1,2,\ldots, N=\mbox{dim}\H\}$ of the system:
\begin{equation}
  \op{\rho} = \sum_{n=1}^N \left[\rho_{nn} \ket{n}\bra{n}
              +\sum_{m>n}\left(\rho_{nm} \ket{n}\bra{m} 
              + \rho_{nm}^* \ket{m}\bra{n} \right) \right]
\end{equation}
such that the diagonal elements $\op{\rho}_{nn}$ in this expansion determine the
populations of the energy eigenstates $\ket{n}$, while the off-diagonal elements
$\rho_{nm}$ for $n \neq m$ determine the coherences between energy eigenstates. 
The latter distinguish coherent superposition states $\ket{\Psi}=\sum_{n=1}^N 
c_n\ket{n}$ from statistical ensembles of energy eigenstates (i.e., mixed states)
$\op{\rho}=\sum_{n=1}^N w_n \ket{n}\bra{n}$.   To see the difference between the
two, consider
$\op{\rho}_1 = \frac{1}{2}\left[\begin{array}{cc} 1 & 0 \\ 0 & 1 \end{array}\right]$
and 
$\op{\rho}_2 = \frac{1}{2}\left[\begin{array}{cc} 1 & 1 \\ 1 & 1 \end{array}\right]$.
$\op{\rho}_1$ is diagonal and represents a two-level system in an incoherent mixed
state with equal populations in states $\ket{1}$ and $\ket{2}$ but no correlation 
between both states.  Note that such a state can\emph{not} be represented by a 
wavefunction.  $\op{\rho}_2$ has off-diagonal elements and diagonalization shows
that it represents the coherent superposition state $\ket{\Psi}=\frac{1}{\sqrt{2}}
[\ket{1}+\ket{2}]$ since we have $\op{\rho}=\ket{\Psi}\bra{\Psi}=\frac{1}{\sqrt{2}}
[1,1]^T\times\frac{1}{\sqrt{2}}[1,1]$.

\section{Dynamics of dissipative quantum control systems}

For Hamiltonian systems the evolution of the state $\op{\rho}(t)$ with 
$\op{\rho}(t_0)=\op{\rho}_0$ is governed by
\begin{equation} \label{eq:rho_evol}
  \op{\rho}(t) = \op{U}(t) \op{\rho}_0 \op{U}(t)^\dagger,
\end{equation}
where $\op{U}(t)$ is the evolution operator satisfying the Schrodinger equation
\begin{equation} \label{eq:SE}
  \rmi\hbar \frac{d}{dt}\op{U}(t) = \op{H}(\vec{f})\op{U}(t), \qquad 
  \op{U}(0)=\op{I},
\end{equation}
and $\op{I}$ is the identity.  $\op{\rho}(t)$ also satisfies the quantum
Liouville equation
\begin{equation} \label{eq:LE}
  \rmi\hbar \frac{d}{dt}\op{\rho}(t) 
  = [\op{H}(\vec{f}),\op{\rho}(t)]
  = \op{H}(\vec{f})\op{\rho}(t) - \op{\rho}(t)\op{H}(\vec{f}).
\end{equation}
$\op{H}(\vec{f})$ is the total system Hamiltonian, which depends on the 
control fields $f_m$:
\begin{equation} \label{eq:H_expansion}
  \op{H}(\vec{f}) = \op{H}_0 + \sum_{m=1}^M f_m(t) \op{H}_m,
\end{equation}
where $\op{H}_0$ is the internal Hamiltonian and $\op{H}_m$ is the interaction 
Hamiltonian for the field $f_m$ for $1\le m\le M$.

The advantage of the Liouville equation (\ref{eq:LE}) over the unitary evolution 
equation (\ref{eq:rho_evol}) is that is can easily be adapted for dissipative 
systems by adding a dissipation (super)operator $\L_D[\op{\rho}(t)]$:
\begin{equation} \label{eq:dLE}
   \rmi\hbar\dot{\rho}(t) = [\op{H}_0,\op{\rho}(t)] +
   \sum_{m=1}^M f_m(t) [\op{H}_m,\op{\rho}(t)] + \rmi\hbar\L_D[\op{\rho}(t)].
\end{equation}
Under certain assumptions (semi-group dynamics, norm continuity and conservation
of probability), it can be shown that the dissipation operator has the form~\cite{1}
\begin{equation} \label{eq:Ld}
  \L_D[\op{\rho}(t)] =
  \frac{1}{2} \sum_s \left( [\op{V}_s\op{\rho}(t),\op{V}_s^\dagger] +
                     [\op{V}_s,\op{\rho}(t)\op{V}_s^\dagger] \right),
\end{equation}
where the $\op{V}_s$ are arbitrary $N\times N$ matrices, i.e., bounded operators
on the Hilbert space $\H$.

\section{Population relaxation and phase decoherence}

In general, uncontrollable interactions of the system with its environment lead
to two types of dissipation: population relaxation (decay) and phase decoherence
(or dephasing).  

The former occurs, for instance, when a quantum particle in state $\ket{n}$ 
spontaneously emits a photon and decays to another quantum state $\ket{k}$, 
which changes the populations according to
\begin{equation} \label{eq:poptrans}
 \dot{\rho}_{nn}(t)
 = -\frac{\rmi}{\hbar}([\op{H}(\vec{f}),\op{\rho}(t)])_{nn}
   +\sum_{k\neq n} \left[\gamma_{nk}\rho_{kk}(t)-\gamma_{kn}\rho_{nn}(t)\right]
\end{equation}
where $\gamma_{kn}\rho_{nn}$ is the population loss for level $\ket{n}$ due to 
transitions $\ket{n}\rightarrow\ket{k}$, and $\gamma_{nk}\rho_{kk}$ is the 
population gain caused by transitions $\ket{k}\rightarrow\ket{n}$.  The total
population relaxation rate $\gamma_{kn}$ is determined by the lifetime of the
state $\ket{n}$, and for multiple decay pathways, the relative probability for
the transition $\ket{n}\rightarrow\ket{k}$.  

The latter occurs when the interaction with the enviroment destroys the phase
correlations between states, which leads to a decay of the off-diagonal elements
of the density matrix: 
\begin{equation} \label{eq:dephasing}
 \dot{\rho}_{kn}(t)
  = -\frac{\rmi}{\hbar}([\op{H}(\vec{f}),\op{\rho}(t)])_{kn}
    -\Gamma_{kn}\rho_{kn}(t)
\end{equation}
where $\Gamma_{kn}$ (for $k\neq n$) is the dephasing rate between $\ket{k}$ and 
$\ket{n}$.  Note that population relaxation always induces dephasing since decay
destroys the phase correlations between states.  However, there may be other 
sources that contribute to the loss of coherence of the system.  

The effects of dephasing and population relaxation can be accounted for by 
adding a dissipation super-operator defined by $(\L_D[\op{\rho}(t)])_{\ell m} 
= (\L_D)_{\ell m,nk} \rho_{nk}(t)$ to the Liouville equation.  The latter can
be represented by an $N^2 \times N^2$ matrix whose non-zero elements are
\begin{equation}
  \begin{array}{l}
  (\L_D)_{kn,kn} = -\Gamma_{kn}, \quad 
  (\L_D)_{nn,kk} = +\gamma_{nk}, \quad  k \neq n \\
  (\L_D)_{nn,nn} = - \sum_{n\neq k} \gamma_{kn}.
 \end{array}
\end{equation}

Population decay and dephasing allow us to overcome kinematical constraints such
as unitary evolution to create statistical ensembles from pure states, and pure 
states from statistical ensembles, which is important for many applications such
as optical pumping.  However, there are instances when this is not desirable such
as in quantum computing, where these effects destroy quantum information.  Hence, 
there are situations when we would like to prevent decay and dephasing.  A cursory
glance at the quantum Liouville equation for coherently driven, dissipative systems
(\ref{eq:dLE}) suggests that it might be possible to prevent population and phase
relaxation by applying suitable control fields such that
\begin{equation}
   \sum_{m=1}^M f_m(t) [\op{H}_m,\op{\rho}(t)]+\rmi\hbar\L_D[\op{\rho}(t)] = 0.
\end{equation}
Unfortunately, however, a more careful analysis reveals that this is \emph{not} 
possible, in general, as we shall now show explicitly for a two-level system, 
or qubit in quantum computing parlance.

\section{Dynamics of a dissipative, coherently driven 2-level system}

The Hamiltonian for a driven two-level system with energy levels $E_1<E_2$ is
\begin{equation}
  \op{H}[\vec{f}(t)]  = \op{H}_0 + f_1(t) \op{H}_1 + f_2(t) \op{H}_2
\end{equation}
where $\op{H}_0$ is the internal Hamiltonian and $\op{H}_1$ and $\op{H}_2$ are
Hamiltonians describing the interaction with independent (real-valued) control
fields $f_1(t)$ and $f_2(t)$.  For a spin system, for example, the two control
fields might be two orthogonal polarizations of an electromagnetic field affecting
rotations about two orthogonal axes.  In general, we can assume without loss of
generality that the internal and interaction Hamiltonians have the form:
\[
  \op{H}_0 =    \left[ \begin{array}{cc} E_1 & 0 \\ 0 & E_2 \end{array} \right], \;
  \op{H}_1 = d_1\left[ \begin{array}{cc} 0 & 1   \\ 1 & 0   \end{array} \right], \;
  \op{H}_2 = d_2\left[ \begin{array}{cc} 0 & -\rmi\\ \rmi & 0 \end{array} \right].
\]
where $d_1$, $d_2$ are the (real-valued) dipole moments for the transition and 
$\omega=(E_2-E_1)/\hbar$ is the transition frequency.

We can re-write the Liouville equation in matrix form in a higher dimensional 
space (Liouville space).  Straightforward computation shows
\begin{equation} \label{eq:dissLiouv1}
    \frac{d}{dt}\lket{\rho(t)}
  = -\rmi [\L_0 + f_1(t) \L_1 + f_2(t) \L_2 + \rmi\L_D] \lket{\rho(t)}
\end{equation}
where $\lket{\rho(t)} = [\rho_{11}(t), \rho_{12}(t), \rho_{21}(t), \rho_{22}(t)]^T$
and the free dissipative evolution is given by 
\begin{equation} \label{eq:Ldiss}
   \L_0 = \omega
   \left( \begin{array}{cccc}
          0 & 0 & 0 & 0 \\
          0 & -1 & 0 & 0 \\
          0 & 0 & +1 & 0 \\
          0 & 0 & 0 & 0
          \end{array} \right), \;
  \L_D =
  \left( \begin{array}{cccc}
          -\gamma_{21} & 0 & 0 & \gamma_{12} \\
          0 & -\Gamma & 0 & 0 \\
          0 & 0 & -\Gamma & 0 \\
          \gamma_{21} & 0 & 0 & -\gamma_{12}
          \end{array} \right)
\end{equation}
with $\gamma_{12}$ being the rate of population relaxation from $\ket{2}$ to 
$\ket{1}$, $\gamma_{21}$ the rate of population relaxation from $\ket{1}$ to 
$\ket{2}$, and $\Gamma$ the dephasing rate.  The control action is given by
\begin{equation}
   \L_1 = \frac{d_1}{\hbar}
   \left( \begin{array}{cccc}
           0  & -1 & +1 & 0 \\
           -1 & 0  &  0 & +1 \\
           +1 & 0  &  0 & -1 \\
           0  & +1 & -1 & 0
          \end{array} \right) , \;
  \L_2 = \frac{d_2}{\hbar}
  \left( \begin{array}{cccc}
          0  & -\rmi & -\rmi & 0 \\
          +\rmi & 0  &  0 & -\rmi \\
          +\rmi & 0  &  0 & -\rmi \\
          0  & +\rmi & +\rmi & 0
         \end{array} \right)
\end{equation}
Observe that the matrix elements of the control Liouville operators $\L_1$ and 
$\L_2$ are zero where the matrix elements of the dissipation operator $\L_D$ are
non-zero, and vice versa.  Thus, no matter how we choose the control fields, we 
cannot cancel the effect of the dissipative terms without introducing additional
terms such as measurements and feedback \cite{2} or the ability to control the 
coupling of the system to the reservoir \cite{3}.

Note that the dissipation operator as defined in (\ref{eq:Ldiss}) is equivalent
to the Lindblad form since inserting
\begin{equation}
  \op{V}_1 = \left(\begin{array}{cc}
              0 & 0 \\ \sqrt{\gamma}_{21} & 0
             \end{array}\right), \quad
  \op{V}_2 = \left(\begin{array}{cc}
              0 & \sqrt{\gamma}_{12} \\ 0 & 0
             \end{array}\right), \quad
  \op{V}_3 = \left(\begin{array}{cc}
              \sqrt{2\tilde{\Gamma}} & 0 \\ 0 & 0
             \end{array}\right)
\end{equation}
with $\tilde{\Gamma}=\Gamma-\frac{1}{2}(\gamma_{12}+\gamma_{21})$ into (\ref{eq:Ld})
gives
\begin{eqnarray*}
 \L_D[\op{\rho}(t)]
  &=& \frac{1}{2} \sum_{s=1}^3 \left( [\op{V}_s\op{\rho}(t),\op{V}_s^\dagger] +
                     [\op{V}_s,\op{\rho}(t)\op{V}_s^\dagger] \right) \\
  &=& \left[ \begin{array}{cc}
             -\gamma_{21} \rho_{11}  & -\frac{1}{2}\gamma_{21} \rho_{12} \\
             -\frac{1}{2}\gamma_{21} \rho_{21} & \gamma_{21} \rho_{11}
             \end{array} \right] 
            + \left[ \begin{array}{cc}
             \gamma_{12} \rho_{22}  & -\frac{1}{2}\gamma_{12} \rho_{12} \\
             -\frac{1}{2}\gamma_{12} \rho_{21} & -\gamma_{12} \rho_{22}
             \end{array} \right] 
           - \left[ \begin{array}{cc}
             0   & \tilde{\Gamma} \rho_{12} \\
             \tilde{\Gamma} \rho_{21} & 0
             \end{array} \right] \\
  &=& \left( \begin{array}{cc}
             -\gamma_{21}\rho_{11}+\gamma_{12}\rho_{22} & -\Gamma\rho_{12}\\
             -\Gamma\rho_{21} & \gamma_{21}\rho_{11}-\gamma_{12}\rho_{22}
             \end{array}\right)
\end{eqnarray*}
which is equivalent to $\L_D\lket{\rho(t)}$ with $\L_D$ as in (\ref{eq:Ld}).

\section{Dissipation and entropy}

One of the main consequences of dissipation is that interactions of the system 
with a bath (environment) can change the entropy and purity of the system.  A 
useful measure of the entropy and purity for our purposes is $1-\Tr[\rho(t)^2]$, 
which essentially determines the Renyi entropy of the system, although the latter
is often defined to be $-\log\Tr[\rho(t)^2]$.  For a non-dissipative, coherently
driven quantum system the entropy is conserved because the evolution must remain
unitary.  Dissipation provides new opportunities for control by enabling us to
reach states outside the orbit of the initial state under the unitary group~\cite{4}, 
especially states whose entropy differs from the initial state.  \emph{Dephasing}, 
for instance, converts coherent superposition states into uncorrelated statistical
mixtures of energy eigenstates and hence enables us, in principle, to convert a 
given pure state into an arbitrary mixed state by creating a superposition state
using coherent control, and letting the coherences decay.  \emph{Population 
relaxation} allows us to convert a high entropy mixed state into a (zero entropy)
pure state and vice versa.  

\subsection{Conversion of a pure state into a mixed state}

When only pure dephasing occurs a coherent superposition of energy eigenstates
$\ket{\Psi}=\sum_{n=1}^N c_n \ket{n}$ with $\sum_{n=1}^N c_n c_n^* =1$ decays 
into a statistical mixture of the states $\ket{n}$ with a discrete probability
distribution $w_n=|c_n|^2$ for $1\le n\le N$.  For dephasing times much greater
than the control time we can design a control field that transforms the initial
state into the required coherent superposition without worrying about dephasing,
and then turn the field off to let dephasing transform this superposition state
into the desired mixed state.  However, if significant dephasing occurs during 
the coherent control phase, either due to rapid dephasing, or because the control
process takes too long, then this approach will fail.   

For instance, consider a system with two non-degenerate energy levels.  Suppose
we wish to transform the pure state $\ket{1}$ into a statistical mixture of the
states $\ket{1}$ and $\ket{2}$.  Based on geometric control for non-dissipative 
systems, we might try to apply a resonant Gaussian control pulse with effective
pulse area $\frac{\pi}{2}$, which would create the superposition state $\ket{\Psi}
=\frac{1}{\sqrt{2}}(\ket{1}+\ket{2})$ in the non-dissipative case, and hope that
decoherence will convert this state into the desired mixed state.  Our control 
calculations indicate, however, that this scheme will fail for dephasing rates
of the order of the Rabi frequency of the control pulse.  However, straight-forward
optimization with respect to the effective pulse area and length of the control 
pulse indicates that the pulse length and pulse area can be chosen such as to 
achieve the desired result.  For instance, by increasing the effective pulse 
area of a Gaussian pulse lasting 50 vibrational periods from the predicted value
of $\frac{\pi}{2}$ for a closed system to $0.81\pi$, we were able to create the
desired maximum entropy state for a dephasing rate $\Gamma=0.1$ in just over 50
vibrational periods.

\subsection{Conversion of a mixed state into a pure state}

A perhaps even more important application of controlled dissipative dynamics in
quantum optics is optical pumping to drive a mixed-state system into a desired 
pure state using a combination of coherent control and population relaxation 
from an excited state.  For instance, suppose we have a cloud of cold atoms 
whose electronic ground state is three-fold degenerate.  If the system is not
prepared in a particular pure state, it will usually be in a mixture of the 
three degenerate substates, which we shall denote by $\ket{1}$, $\ket{2}$ and
$\ket{3}$ for simplicity.  For many applications, e.g., in quantum computing, 
it is crucial to prepare the system in a certain pure initial state.  As we 
have seen, this is an aim generally impossible to realize by coherent control
alone.  To be able to take advantage of spontaneous emission to increase the 
purity of the system, we must couple the ground state to an excited electronic
state with a finite lifetime.  The sublevels of the ground and excited states
can be coupled in various ways depending on the polarization of the field.  
The trick is to select the right coupling.

\begin{figure}
\begin{center}
\input{figures/figure3.latex}
\end{center}
\caption{Optical pumping for a degenerate two-level system.  Transition diagrams
for the control field and population relaxation}\label{fig3a}
\end{figure}
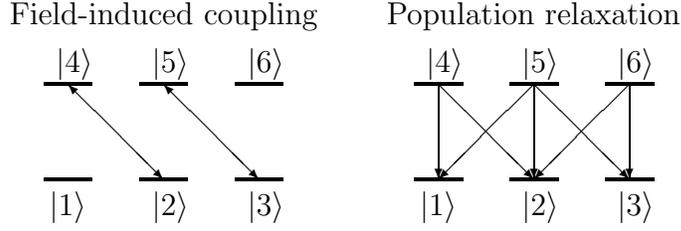

\begin{figure}
\begin{center}
\myincludegraphics[width=0.49\textwidth]{figures/pdf/figure3a.pdf}{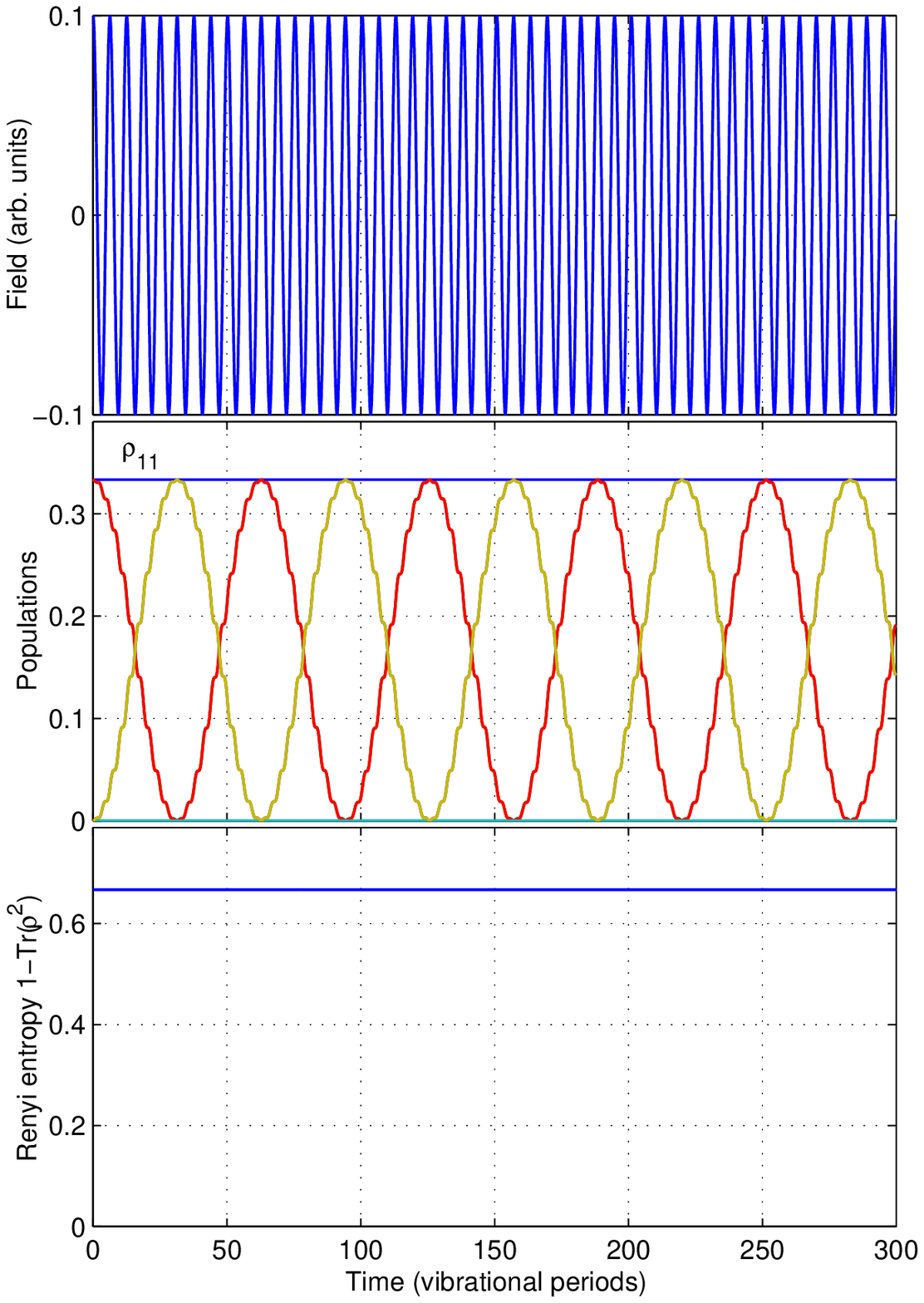}
\myincludegraphics[width=0.49\textwidth]{figures/pdf/figure3b.pdf}{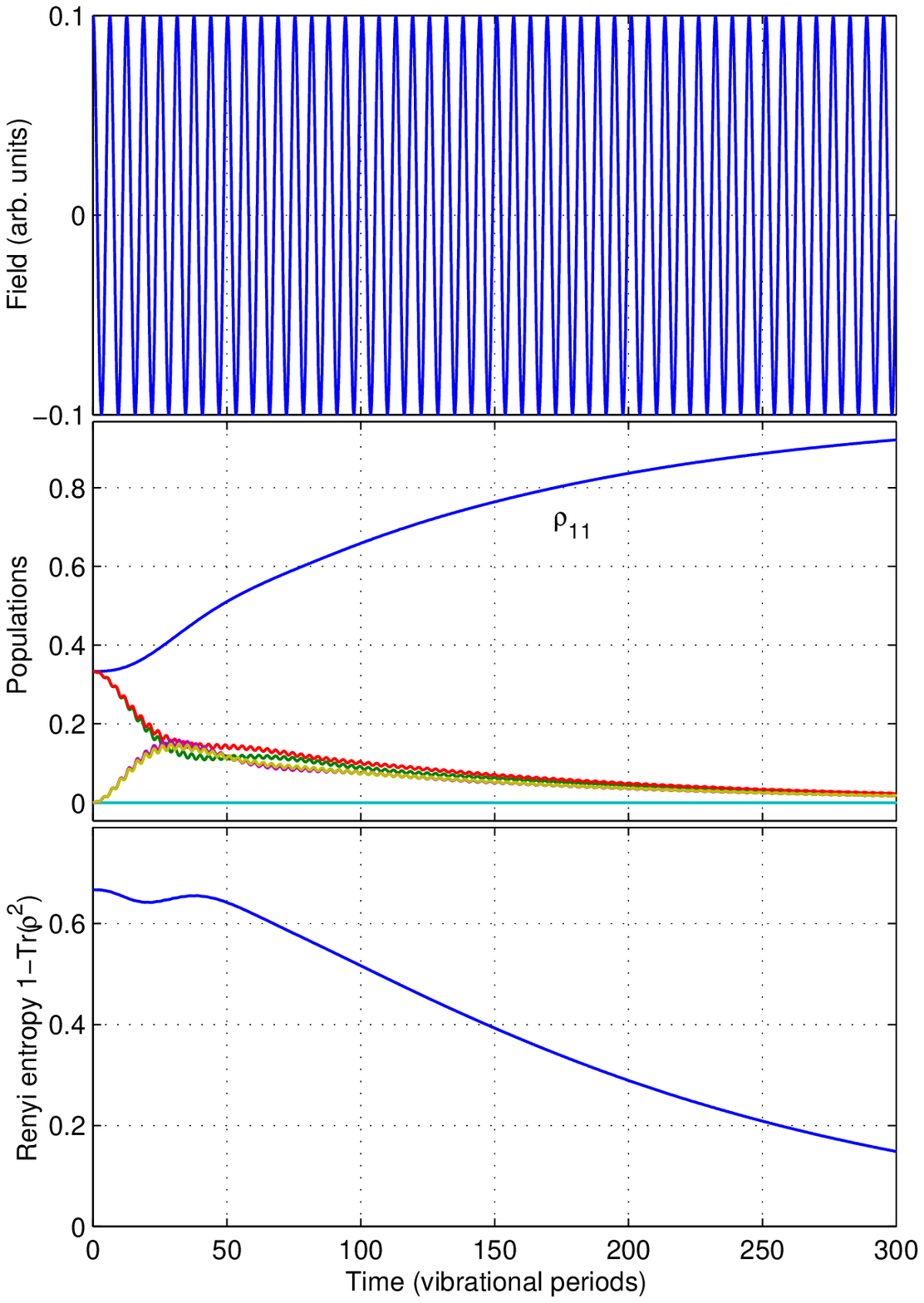}
\end{center}
\caption{Optical pumping for a degenerate two-level system.  Without population
relaxation, the coherent control field induces population oscillations between
states $\ket{2}$ and $\ket{5}$ as well as $\ket{3}$ and $\ket{6}$ (Rabi
oscillations) and the entropy of the system remains constant (left).  Population
relaxation dramatically changes the effect of the control field, leading to an
effective pumping of population into the lower left sublevel $\ket{1}$, and an
entropy reduction as the system approaches a pure state (right).}\label{fig3b}
\end{figure}

For example, suppose the upper level is also three-fold degenerate and the
coupling induced by the control field is as indicated in figure \ref{fig3a}, 
i.e., the field couples states $\ket{2}$ and $\ket{5}$, as well as $\ket{3}$
and $\ket{6}$.  The excited states can emit a photon and return to one of 
the ground states.  Certain transitions are prohibited by atomic selection
rules; the allowed decay modes are indicated in figure \ref{fig3a} (right).  
The simplest optical pumping schemes involve applying a constant amplitude 
field resonant with the transition frequency between the two levels and 
suitably polarized to couple only the levels indicated in figure \ref{fig3a}.  
Without population relaxation due to spontaneous emission, the field merely 
leads to population cycling between states $\ket{2}$, $\ket{5}$, and $\ket{3}$,
$\ket{6}$, respectively.  Adding population relaxation changes the effect of
the control field dramatically, leading to an accumulation of the population
in state $\ket{1}$ as figure \ref{fig3b} shows.  If the control field is 
applied for a sufficiently long time, all the population will eventually
accumulate in state $\ket{1}$.

In the previous optical pumping scheme a simple constant amplitude resonant 
control field was sufficient to achieve the objective of driving the system 
into the desired pure state.  However, this is not always the case.  Some
applications of optical pumping such as laser cooling of internal molecular
degrees of freedom rely on the interplay of carefully selected control pulses
and dissipation.  For example, a molecular vapor at room temperature consists
of a statistical mixture of molecules in many different ro-vibrational states.
Due to many closely spaced energy levels and lack of selection rules, there 
are many possible transitions with various transition probabilities that can
be excited by applying a control field, and many different decay pathways.  
The situation is further complicated by the fact that the timescales for 
coherent control and population relaxation are often quite different.  The 
problem thus appears to be nearly hopeless.  Yet, it has been shown that 
this problem can be addressed successfully using optimal control for 
dissipative systems and creative control strategies \cite{5}.

An approach that is especially promising for systems where the timescales 
for control and dissipation are quite different (as in our molecular cooling
problem) involves breaking the problem up into a sequence of excitation and
relaxation steps.  The goal in each step is to use control theory to design
control fields to transfer the system from its initial state to a kinematically
equivalent, dynamically reachable state, which has the same entropy but is
likely to decay into a state with lower entropy.  In principle, the entropy
of the system can be decreased until it is zero and the system is the desired
pure state.  The main difficulty of this approach is the choice of suitable 
target states for each optimization step, which requires a good understanding
of the effects of population relaxation and dephasing on various kinematically
equivalent states, in order to assure that the selected states will decay 
into a lower entropy state.

\section{Conclusion}

We have shown that although coherent control of the system dynamics in absence
of feedback or the ability to alter the interaction of the system with its
environment cannot compensate for dissipative effects such as dephasing and 
population relaxation in open systems, such effects need not be detrimental to
control, but are in fact crucial for many interesting applications such as
ensemble preparation or system purification by laser cooling.

\section*{Acknowledgements}
SGS would like to thank the Cambridge-MIT Institute for financial support.

\end{document}

%% file: figures/figure3.latex
\setlength{\unitlength}{3947sp}%
\begingroup\makeatletter\ifx\SetFigFont\undefined%
\gdef\SetFigFont#1#2#3#4#5{%
  \reset@font\fontsize{#1}{#2pt}%
  \fontfamily{#3}\fontseries{#4}\fontshape{#5}%
  \selectfont}%
\fi\endgroup%
\begin{picture}(4033,1344)(233,-661)
\thicklines
\put(1576,164){\line( 1, 0){300}}
\put(376,164){\line( 1, 0){300}}
\put(976,164){\line( 1, 0){300}}
\put(976,-436){\line( 1, 0){300}}
\put(376,-436){\line( 1, 0){300}}
\put(1576,-436){\line( 1, 0){300}}
\thinlines
\put(526,164){\vector(-1, 1){  0}}
\put(526,164){\vector( 1,-1){600}}
\put(1126,164){\vector(-1, 1){  0}}
\put(1126,164){\vector( 1,-1){600}}
\put(1651,239){\makebox(0,0)[lb]{\smash{\SetFigFont{12}{14.4}{\familydefault}{\mddefault}{\updefault}$|6\rangle$}}}
\put(526,-661){\makebox(0,0)[b]{\smash{\SetFigFont{12}{14.4}{\familydefault}{\mddefault}{\updefault}$|1\rangle$}}}
\put(1051,-661){\makebox(0,0)[lb]{\smash{\SetFigFont{12}{14.4}{\familydefault}{\mddefault}{\updefault}$|2\rangle$}}}
\put(1651,-661){\makebox(0,0)[lb]{\smash{\SetFigFont{12}{14.4}{\familydefault}{\mddefault}{\updefault}$|3\rangle$}}}
\put(451,239){\makebox(0,0)[lb]{\smash{\SetFigFont{12}{14.4}{\familydefault}{\mddefault}{\updefault}$|4\rangle$}}}
\put(1051,239){\makebox(0,0)[lb]{\smash{\SetFigFont{12}{14.4}{\familydefault}{\mddefault}{\updefault}$|5\rangle$}}}
\thicklines
\put(3901,164){\line( 1, 0){300}}
\put(2701,164){\line( 1, 0){300}}
\put(3301,164){\line( 1, 0){300}}
\put(3301,-436){\line( 1, 0){300}}
\put(2701,-436){\line( 1, 0){300}}
\put(3901,-436){\line( 1, 0){300}}
\thinlines
\put(2851,164){\vector( 1,-1){600}}
\put(3451,164){\vector( 1,-1){600}}
\put(3451,164){\vector( 0,-1){600}}
\put(4051,164){\vector( 0,-1){600}}
\put(4051,164){\vector(-1,-1){600}}
\put(3451,164){\vector(-1,-1){600}}
\put(2851,164){\vector( 0,-1){600}}
\put(3976,239){\makebox(0,0)[lb]{\smash{\SetFigFont{12}{14.4}{\familydefault}{\mddefault}{\updefault}$|6\rangle$}}}
\put(2851,-661){\makebox(0,0)[b]{\smash{\SetFigFont{12}{14.4}{\familydefault}{\mddefault}{\updefault}$|1\rangle$}}}
\put(3376,-661){\makebox(0,0)[lb]{\smash{\SetFigFont{12}{14.4}{\familydefault}{\mddefault}{\updefault}$|2\rangle$}}}
\put(3976,-661){\makebox(0,0)[lb]{\smash{\SetFigFont{12}{14.4}{\familydefault}{\mddefault}{\updefault}$|3\rangle$}}}
\put(2776,239){\makebox(0,0)[lb]{\smash{\SetFigFont{12}{14.4}{\familydefault}{\mddefault}{\updefault}$|4\rangle$}}}
\put(3376,239){\makebox(0,0)[lb]{\smash{\SetFigFont{12}{14.4}{\familydefault}{\mddefault}{\updefault}$|5\rangle$}}}
\put(1126,539){\makebox(0,0)[b]{\smash{\SetFigFont{12}{14.4}{\rmdefault}{\mddefault}{\updefault}Field-induced coupling}}}
\put(3451,539){\makebox(0,0)[b]{\smash{\SetFigFont{12}{14.4}{\rmdefault}{\mddefault}{\updefault}Population relaxation}}}
\end{picture}